\documentclass[a4paper, 10pt, conference]{IEEEtran}

\usepackage{graphicx}
\usepackage[colorlinks=true,hyperfootnotes=true,breaklinks=true]{hyperref}
\usepackage[T1]{fontenc}
\usepackage{amsmath}
\usepackage{psfrag}
\usepackage{auto-pst-pdf}
\usepackage{harvard}
\usepackage{balance}

\author{Krzysztof Ku{\l}akowski, Krzysztof Malarz, Ma{\l}gorzata J. Krawczyk \\
AGH University of Science and Technology \\
Faculty of Physics and Applied Computer Science \\
30-059, Krakow, Poland \\
Email: kulakowski@fis.agh.edu.pl}

\title{\textbf{HEAVY CONTEXT DEPENDENCE---DECISIONS OF UNDERGROUND SOLDIERS}}
\begin{document}
\maketitle




\section*{\textbf{KEYWORDS}}

History of WWII; Decision trees; Minimax; Percolation

\section*{\textbf{ABSTRACT}}

An attempt is made to simulate the disclosure of underground soldiers in terms of theory of networks. The coupling mechanism between the network nodes is the possibility that a disclosed soldier is going to disclose also his acquaintances. We calculate the fraction of disclosed soldiers as dependent on the fraction of those who, once disclosed, reveal also their colleagues. The simulation is immersed in the historical context of the Polish Home Army under the communist rule in 1946-49.

\section*{\textbf{INTRODUCTION}}
                  
In the long-lasting discussion about how predictions are or may be possible in social sciences \cite{1,2,3} scholars agree that the conditions of a successful prediction are tough and rarely satisfied.
In \cite{4}, the following warning has been formulated by Karl Popper: `{\it science can predict the future only if the future is predetermined---if, as it were, the future is present in the past, telescoped in it}'.
This means in particular, that the system to be predicted should remain stationary and isolated from external factors which could destroy the accuracy of the intended prediction.
As stated by Popper, these requirements are not met by the modern society.
In more recent writings and presentations \cite{2,5}, Bruce Edmonds indicates that predictive models of complex (as social) systems are rare, and that models which only pretend to be predictive can actually be dangerous in untrained hands of stakeholders.
In the conclusions of an influential book on predictions \cite{3}, Nate Silver writes: `{\it There is no reason to conclude that the affairs of men are becoming more predictable. (...) The same sciences that uncover the laws of nature are making the organization of society more complex}'.
The power of the Silver's method seems to be a very careful data analysis for issues where a prediction is possible, leaving other issues aside, and his mastery is to distinguish these two kinds.
Adherents of social simulations often provide sets of rules, which have to be complied with to get a credible outcome \cite{2,6,7}. When faced with these warnings and caveats, it is tempting to expand the list of examples, yet short, where a prediction is possible. 

We are going to discuss a situation when the complexity of decisions is limited by the pressure of an external power. Once basic human needs are threated, one can expect that main interest and activity will be directed to minimize the threat. In particular, when an individual's safety is at risk in conditions of limited information, the strategy of minimizing the maximal loss (minimax) is the most straightforward solution. In our case, this strategy was equivalent to an active collaboration with the regime. This is our main assumption here. On the contrary to many social phenomena, here the context makes the problem simpler.

We propose to consider a historical example: the social network of soldiers of the Home Army (Armia Krajowa, AK) in Poland, conscripted to regular army under Soviet command in late 40's. In this network, node X is connected to node Y (a link from X to Y) if X knows the partisan activity of Y. Accordingly, X can inform against Y to the communist security service. The problem of the fraction of soldiers who will be revealed is discussed here in the frames of the percolation theory on networks \cite{8,9,10}. Although these frames are clearly more wide, than the particular issue taken from history of Poland, we concentrate on this example to include the cognitive and social contexts \cite{11} to the model. In our case, these are: the Stalinist reality encoded in norms and expectations, and the relations of power. Although the generality of the model approach is narrowed in this way, these contexts allow to fix and justify the model assumptions.

The aim of this work is an analysis of the time evolution of the political attitudes of the soldiers, with particular attention to the conformist behavior of some of them with respect to the communist power. Although this problem is precisely located in the past history, it is relevant for similar phenomena, both those in more remote and those in more contemporary times. In our case, we are going to trace the path from a partisan of the Polish Underground State to an informer of the communist secret police.

Our method includes a reconstruction of personal power networks (PPN) of the soldiers, based on writings of Polish historians \cite{12,13,14}. This step is motivated by the concept of figurations, as introduced by Norbert Elias \cite{15}.
By a personal power network of node X we mean a set of ties to X from persons who have some kind of power over X.
Links in these networks mean the relations of power of different kinds, from a threat of life to a referent power including, in a broader sense, family ties. Subsequent decisions of the soldiers are described in terms of the decision trees \cite{16}, with information---at their time---incomplete.

In the next two sections we make an attempt to highlight the context. Further, we try to reconstruct the personal power networks of a soldier, and we propose a scheme of questions
which should be posed to know if a respondent is going to disclose himself as an ex-partisan, to join the communist party and to inform about his colleagues. This scheme is presented as a decision tree. We analyse briefly what could be the answers, provided that the respondent tries to minimize his maximal loss and that he tells true. The coupling of the soldiers' decisions makes a space for a simulation, performed in terms of theory of networks. Final conclusions close the text.

\section*{\textbf{THE COGNITIVE CONTEXT: THE STALINIST REALITY}}

In 1943, the underground AK in Poland had about 200 thousands of soldiers \cite{12}. Fighting against the Germans, some AK groups happened to cooperate with the Red Army; yet, as the front moved west, these groups were disarmed by the People's Commissariat for Internal Affairs (NKVD). The AK privates have been directed to the Polish military units (WP) fighting against the Germans under Soviet command, the AK officers have been arrested. Military actions of AK against the Red Army has been ruled out  by the AK headquarters. In May 1944, the Soviet authorities ordered a general mobilization of all Polish men between 18 and 55 years old. In July 1944, Polish communist security forces of about 20 thousands have been created by NKVD. Both these bodies concentrated on chasing after AK soldiers; between July and December 1944, about 30 thousands of them were arrested. New penal code of WP allowed to try civilians in political cases, with the death penalty for having a radio or for not informing the communist 
authorities about what what the latters considered as `criminal' actions.

In July 1945, an amnesty was declared, which however did not apply to commanders who did not disclosed themselves and their subordinates. Some AK high-rank commanders accepted the amnesty conditions and asked others to disclose. In June 1946, a referendum was carried out to prove national support for the communists, and the results were falsified by them. In January 1947, national elections have been organized, with simultaneous terror, arrests and murders of political opponents. One month later, the second amnesty released about 25 thousands of prisoners sentenced for less than 5 years. Those who disclosed  themselves were promised to set free. In summer 1948, the disclosed soldiers have been arrested again, many of them subjected to torture during investigations.

Parallel to these processes, the communist policy evolved from the coalition of all `non-fascist' parties in 1945, with AK excluded as a fascist one, to a split and gradual elimination of the agrarian Christian democrats (PSL) and the socialists (PPS). Finally a split of the communist party itself (PPR) appeared, and its moderate faction has been deprived of power in 1948. The remains of PPS were absorbed by PPR in 1948.

\section*{\textbf{THE SOCIAL CONTEXT: THE RELATIONS OF POWER}}

In each army unit in Poland, the counter-intelligence objectives are performed mainly by the security officer, appointed to protect the unit (obiektowy oficer informacji, OOI); this role has not changed since 1945 \cite{13}. Everywhere in the world, the task of counter-intelligence officers is to know everything what could be important for the security of actions. In WP in the 40's, this role was understood as to eliminate `the hostile element'. Between October 1944 and October 1946, soldiers of AK were qualified to this category by definition; the same practice was renewed after October 1947.
Persons in this category have been subjected to a continuous investigation, their promotions have been blocked or delayed, and often they were dismissed from the army.

OOI has been responsible for the control of political conduct of all soldiers, except the unit commander.
Yet, although OOI's have been obliged to inform the unit commander about his results, the content of this information had to be arranged within the political section, a kind of a state in a state. It was the duty of OOI's to collect a set of unofficial informers (nieoficjalny wsp\'o{\l}pracownik, NW).
Basically, the network of NW's could not contain loops; NW's did not know each other, then nobody knew if his colleague or interlocutor was going to denounce him to communist authorities.
OOI's have been controlled precisely, and they had to carry out a detailed documentation of all their actions.
OOI's  have been assessed; an inefficient officer could be dismissed or even sentenced to imprisonment \cite{13}.
Their work was also accounted for by PPR; a refusal to join this party automatically qualified an OOI as a hostile element. Therefore, the fraction of OOI's who were PPR members increased from 51 in 
December 1946 to 88 in December 1948. We note also that central command of the information service has been controlled by the officers of counter-intelligence agencies of the Red Army (SMERSH).

These facts indicate, that OOI's have been played a main role in the personal power networks of our set of AK soldiers in WP. Their power---the ability to influence soldiers' actions---was greater than the power of the commanders, as the rules of the latter were commonly known and limited by detailed military regulations.
 
\section*{\textbf{PERSONAL POWER NETWORKS}}

A PPN for a soldier of AK can be reconstructed only as a standard one; we assume that the variations of PPN's from one soldier to another can be taken into account only in the statistical sense, as explained below. The list of ties in such PPN of a soldier includes his direct commander and his OOI, but also his subordinates, his family and---last but not least--- his acquaintances from AK. Although contacts with the last group have been certainly less vivid, the awareness about  a possible threat of their disclosure was certainly unrelenting. Actually, these latter ties are the only essential ones for the dynamics of our system. The ties with the family can be qualified as the referent power \cite{17}; they have provided a motivation to live, to protect these nears and dears, to plan the future, to make a career. On the other hand, the ties with the commander and even more with OOI have ensured that once a soldier broke the communist rule or would be denounced, his career, his freedom and perhaps even his life are finished.
(Here there is no need to specify the details of fates of those accused and condemned.)
Both these kinds of ties do not constitute independent variables, but rather they should be treated as the context.
The ties with the subordinates in WP can be roughly qualified into the two above contextual kinds. The actual evolution of the system takes place between the soldier and his colleagues from AK; who is going to disclose? whom he will denounce?

\begin{figure}
\begin{center}
\includegraphics[width=.6\columnwidth]{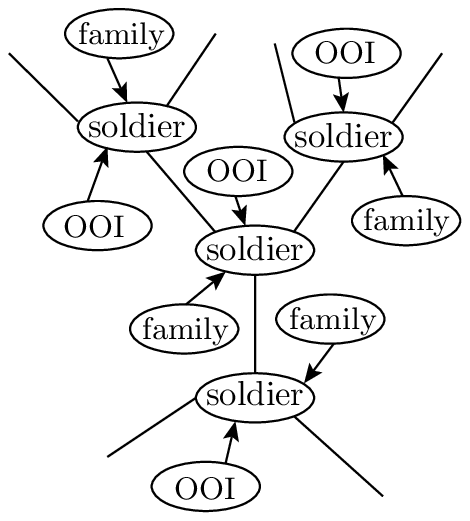} 
\end{center}
\caption{\label{fig-1} A personal power network (PPN) of a soldier. Directed links mark the contextual influence of the soldier's neighbors, family and communist security services (OOI). The link from the subordinates is omitted for clarity of the picture. The links from/to neighbor soldiers are left undirected, as their bidirectional character is important here. They are of dynamic character in the sense that we are interested in the soldier's decisions and how the consequence of these decisions spread along the network. On the contrary, the roles of the family and of the OOI are fixed.}  
\end{figure}

A sketch of the resulting PPN is shown in Fig.~\ref{fig-1}.
The structure of the relevant links between the soldiers of AK remains unknown.
One could imagine that in 1943, this structure was yet closer to a hierarchical one, as such was the organizational structure of underground AK structure.
However, between 1946 and 1950 many high-rank AK commanders have been already disclosed.
We believe that it is more appropriate to assume that the network topology is close to an uncorrelated sparse random network, as the Erd\H{o}s--R\'enyi network \cite{10}. The arguments are as follows: {\it i)} in an underground army, soldiers have been hidden under pseudonyms, what makes the network of eventual mutual disclosures much more sparse and less clustered;
{\it ii)} for the sake of safety, in this network hubs did not appear; high-rank officers known from their legal activities before WWII have been quickly arrested or they managed to emigrate. Here we add that in a more clustered network, the process of disclosure is less likely to stop at the tree-like branch, where an agent A may appear. Therefore, in a more clustered network the process is even more abrupt. Yet to verify the hypothesis on the network structure, an extensive analysis of tens thousands of dossiers would be necessary, what remains out of the frames of this paper.

\section*{\textbf{THE DECISION TREE}}

The task of the construction of a decision tree includes the formulation of the decision criteria: questions which should be asked as to distinguish between those who are going to make different decisions. Although the method is termed `qualitative' in \cite{16}, its accuracy can be evaluated in numbers; this is the fraction of those respondents who make the decision in agreement with what has been predicted according to their answers for the questions.

In our case, the standard method of an evaluation of these questions (a check of the respondents' decisions) is very difficult because the lag of time---even if soldiers of AK are still alive, both their answers and their memories have been transformed by later events. Moreover, even if the research is conducted in late 40's, it would be likely that an untrustworthy pollster would get a bullet through his head. Indeed, when a truthful answer threatens the respondent's life, it is hard to imagine an objective research. Yet, the same concerns several groups in the contemporary world; to allude to these groups is our motivation here.

\begin{figure}
\begin{center}
\includegraphics[width=.9\columnwidth]{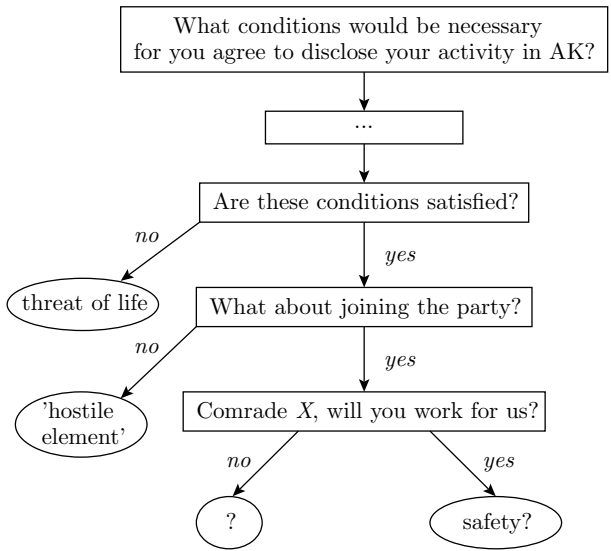} 
\end{center}
\caption{\label{fig-2}The decision tree of a soldier.
Subsequent levels mark the sequence of questions and roles.
The first question is descriptive; the dotted box contains the answer with the description of a personal context.
Although the role of this box is only intermediate in the soldier's decisions, it is placed in the tree to mark that the dialogue could not be limited to the question about disclosure.
The outcomes are marked as ellipses.
`Hostile element' means being qualified as such.
The outcomes could not be expressed as payoffs, because of the lack of information.}
\end{figure}

According to \cite{16}, the groups to be inquired as to form the proper questions should include persons who answer and decide in different ways. Our method here is to replace respondents by documents; then we use both letters between AK authorities and communist officials \cite{18}, written of necessity in an official language, and descriptions \cite{14} of actions of partisans written without any censorship. Even a cursory reading reveals a deep split between the ideas in writings and even the languages of documents which represent the two rival political orientations, and this split is perhaps the most important characteristics of Polish political life at that time \cite{12}. Despite this, our questions should be formulated in a language which could be accepted by a respondent who could give virtually any answer. Yet, two points make our task feasible: {\it i}) the logic structure of the decision tree is based on the answers `yes' or `no', {\it ii}) the answer `no' at a given level 
excludes the 
respondent from the further investigation. The proposed decision tree are shown in Fig.~\ref{fig-2}.

The relations with colleagues from AK can serve to highlight the collective character of the process in the first two levels of decisions: to disclose or not and to join the communist party (PPR or its successor since December 1948, the Polish United Workers' Party) or not.  This collective character is related to the correlations between the fates of soldiers who know each other; once X is investigated, it is likely that he informs about his acquantances Y and Z, those in turn inform about their neighbors etc.
We should note that in the documents signed by the AK authorities \cite{18}, the conditions of the amnesty of 1945 do not contain any demand of revealing other persons.
Despite that, the legal regulation demanded that a soldier revealed all his subordinates \cite{19}. Also, during any trial the policy of the communist functionaries was to reveal officers \cite{12}. As a consequence, many disclosures snowballed. In principle, this kind of processes can be described in terms of the percolation theory \cite{8}. 
In a nutshell, consider the probability $p$ that one disclosed soldier informs about the other one.
If $p$ is larger than some critical value $p_c$, a large part of the network will be disclosed. The actual value of $p_c$ depends on the network topology; in uncorrelated networks, as is the Erd\H{o}s--R\'enyi one, $p_c=\langle k\rangle/(\langle k^2\rangle-\langle k\rangle)$, where $\langle k\rangle$ ($\langle k^2\rangle$) is the mean (squared) number of neighbors of a soldier in the network \cite{10}. The link to the percolation leads to the analogy with modeling epidemics in networks \cite{ken}. Another link to simulations is the equivalence of our problem to the $N$-person Prisoner's Dilemma \cite{20}, which has been investigated in various variants, e.g. \cite{21}.

\section*{\textbf{THE OUTCOMES OF DECISIONS}}

As an outcome of the particular decisions, the related payoffs should be determined, and the decisions should be evaluated by a comparison of the payoffs. However, in our case this procedure cannot be applied for two reasons. First is that the information of our model decision-maker is so limited that even the probabilities of particular events cannot be evaluated by himself. Hence, the same applies to his payoffs. Second is that these decisions are made in the heavy condition of life threat, and therefore the related emotions cannot be excluded.  It is likely that the strategy of a soldier is to minimize his maximal loss \cite{20}.

To remain undisclosed was equivalent to suffer a continuous fear that the partisan activity during the war would be revealed. The risk was high: to be imprisoned, perhaps murdered, with tough consequences also for the family. On the other hand, those who disclosed themselves had to resist an ostracism of those who did not \cite{18}. Yet, the needs of respect and belongingness are less fundamental than the need of safety \cite{22}, and the former two are activated only if the latter is satisfied.

Once a soldier was disclosed, his relations with his former colleagues had been weakened, and gradually lose value. In his plans for the future looking forward to WWIII, common at that time in Poland \cite{23}, was replaced by a vision of a coexistence with the communist power and institutions. Also, in official media the gain control of PPR over PPS has been presented as a fusion of two equivalent parties; we note that PPS had been more socially accepted. Hence the membership of both parties increased much at that time (in thousands: in December 1944 PPR 34, PPS 5; in December 1947 PPR 820, PPS 713) \cite{12}. Those who remained non-party could expect to be passed over in promotions, of delays in being allotted a flat etc. When trying to remove their labels as `a hostile element', to join the ruling party seemed a most straightforward way.

Finally, in the late 40's there was no barriers for the activity of OOI's in WP \cite{24}; it is only after the thaw in 1956 that the drawing of party members into the role of informers has been limited. Consider an ex-AK soldier, with a dossier incriminated in eyes of the ruling power, now a member of the communist party, untrustworthy both for his colleagues in WP (as nobody was trustworthy there) and for his former social environment. What are his {\it cons} and {\it pros} if he is asked, more or less arbitrarily, by an OOI of his unit, to contact from time to time? How could he evaluate the risk of consequences when his answer is `no'? With `yes', he could count on at least neutrality of the apparently omnipotent counter-intelligence services. On the other hand, he could hope that his new role as NW would remain in secret. Many refused, but not all of them. In December 1946, the proportion of NW's in the officer corps of WP was 1:12. In subsequent years, this proportion has systematically increased 
till 
1:5 in December 1949 \cite{13}.  We have no data on the fraction of the `hostile elements' in this group. However, according to the order from 9 December 1948, each NW qualified as a hostile element had himself to be subject of an observation \cite{13}. This indicates, that the number of those people was at least noticeable.

\section*{\textbf{THE SIMULATION}}

The interdependence of the probabilities of disclosing two soldiers connected in the network makes place for an application of statistical mechanics and simulations. If the whole effect can be reduced to a direct influence of one soldier to another, then the fraction of the disclosed soldiers can be found by a direct application of the percolation theory, as a size of the largest connected cluster \cite{8}. However, we have also another mechanism: a soldier could disclose himself without any relations to his neighbour colleagues. These people activate further avalanches of disclosing soldiers, which spread along the network ties. The proportions of these two mechanisms remains unknown. What we can demonstrate is the joint consequence of both of them, with their efficiencies as the parameters of the calculation, for the network topology assumed to be of Erd\H{o}s--R\'enyi type \cite{25}. The results are to be taken as an illustration of the discussed effects rather than a quantitative evaluation. 
This kind 
of interpretation remains in accordance with the famous statement of Richard W. Hamming: `{\it The purpose of computing is insight, not numbers}' \cite{26}. We note on the margin that the question most important for an individual soldier: `what is the probability that I will be arrested', has only recently been recognized as a NP-hard problem (formulated in terms of epidemics) \cite{sha}.

To be more specific, we assume that in our model network, there are three kinds of soldiers. Soldiers of type A do not disclose themselves; also, if they are disclosed, they do not reveal what they know about their neighbours. Soldiers of type B do not disclose themselves, but---once disclosed---they inform about their neighbours. Soldiers of type C disclose themselves and inform about their neighbors. These three types are initially distributed randomly in the Erd\H{o}s--R\'enyi network of $N$ nodes, with their numbers $N_A$, $N_B$ and $N_C$, and their fractions $n_A\equiv N_A/N$, $n_B\equiv N_B/N$ and $n_C\equiv N_C/N$, respectively.
Given $N_C$, we ask how many soldiers will be disclosed?

\begin{figure}
\psfrag{nD}{{\small $n_D$}}
\psfrag{y}{{\small $y$}}
\psfrag{nB+nC}{{\small $n_B+n_C=$}}
\includegraphics[width=\columnwidth]{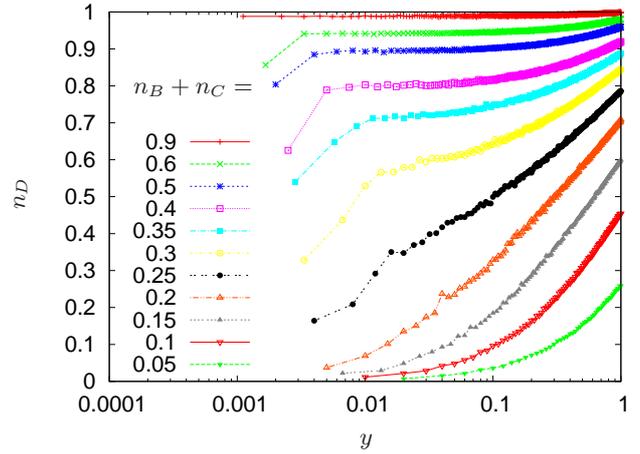}
\caption{\label{fig-3} The fraction $N_D/N \equiv n_D$ of the soldiers in the network who are disclosed as dependent on the fraction $y\equiv N_C/(N_C+N_B)$ of those who disclose (themselves and their neighbours) spontaneously ($N_C$) in the set of those ($N_B+N_C$) who disclose themselves and their neighbors, both spomtaneously and when disclosed. The curves are related to increasing $N_B$ from bottom to top. The simulation is for $N=1000$ nodes with mean degree $\langle k\rangle=5$, and the results are averaged over $N_{\text{run}} = 100$ runs.}
\end{figure}

\begin{figure}
\psfrag{nD}{{\small $n_D$}}
\psfrag{nB+nC}{{\small $n_B+n_C$}}
\psfrag{N, Nrun}{{\small $N$, $N_{\text{run}}=$}}
\includegraphics[width=\columnwidth]{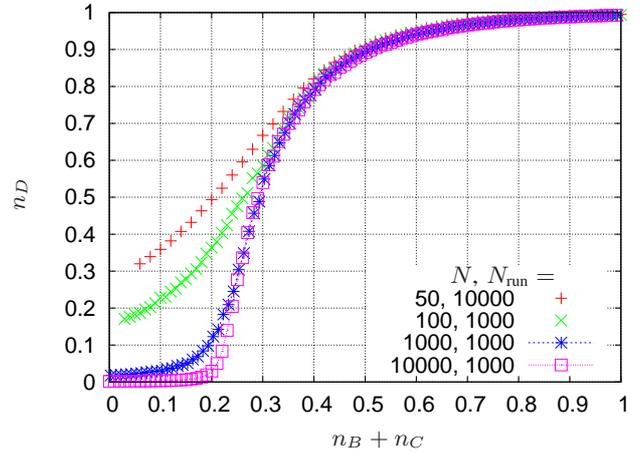}
\caption{\label{fig-4}  The fraction of the soldiers in the network who are disclosed as dependent on $n_B+n_C=(N_B+N_C)/N$, for given (and small) value of $N_C=3$. As we see, for $N_B/N$ large enough, almost the whole network is disclosed even for small values of $N_C$. The transition becomes more sharp for larger networks.}
\end{figure}

\begin{figure}
\psfrag{t}{{\small $t$}}
\psfrag{nB}{{\small $n_B\approx$}}
\psfrag{nD}{{\small $n_D$}}
\includegraphics[width=\columnwidth]{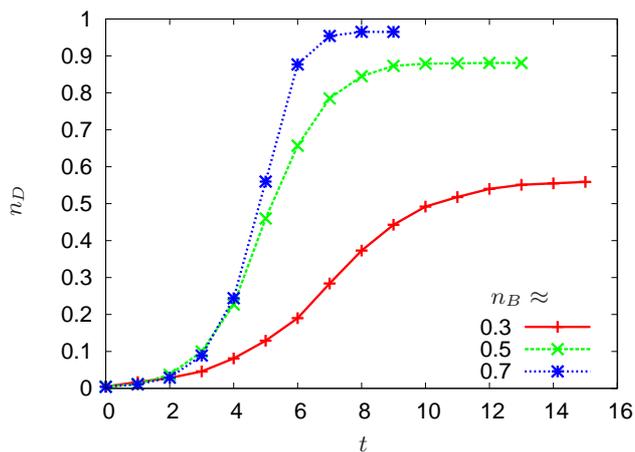}
\caption{\label{fig-5}  Three exemplary curves of the time dependence of the fraction $n_D\equiv N_D/N$ of the soldiers in the network who are disclosed for $N_C=1$, for three values of $(N_B+N_C)/N \approx n_B$ = 0.3, 0.5 and 0.7. The network size is $N$ = 1000.}
\end{figure}

The results are shown in Figs.~\ref{fig-3}-\ref{fig-5}.
Fig.~\ref{fig-3} provides the overall data: the fraction $n_D\equiv N_D/N$ of disclosed soldiers in the whole population.
In Fig.~\ref{fig-4}, we show the same proportion $N_D/N$ as dependent on the ratio $(N_C+N_B)/N$, with an assumption that $N_C$ is a small number; here we put $N_C=3$. The results indicate, that if the attitude of type B is frequent, a small number of `spontaneous disclosers' leads to the disclosure of the practically whole network.
Note that the results for larger networks are more representative, because the statistics is better. As we see, once the percolation threshold (equal to 0.2 for the mean degree $\langle k\rangle=5$) is exceeded, the results increase very quickly, before they stabilize for longer times. In Fig.~\ref{fig-5}, we show three trial examples of the time dependence of the number of soldiers revealed to communist officials.

\section*{\textbf{DISCUSSION}}

To summarize, we argued that a reasonable agent should {\it i)} disclose himself, {\it ii)} join the communist party, {\it iii)} accept the role of NW. We could add one point more: try to defect to the West, but this option is not discussed here. The results of our simulation indicate, that once these actions have been made in social scale, only a small part of the network of the soldiers could remain uncontrolled by the communist power. This is a consequence of the `small-world' property of social networks \cite{25}. The processes described above have been understandable and they should be taken into account, when evaluating the dynamics of social choices in Poland between 1945 and 1949. Moreover, an interplay of incomplete information and of threats of basic human needs is a common component of life in (too) many current societies. The strategy of minimizing the maximal loss seems to provide a key to understand decisions of their members.

Two more general afterthoughts related to predictions of social phenomena seem to be appropriate here.  First is that to predict what people will do, it is highly recommended to contact them and try to understand the way they think. This is not a sufficient condition, but in some circumstances the method can be efficient. As we already noted after Christina H. Gladwin, the accuracy of prediction can be evaluated quantitatively. It is somewhat strange that the method is termed as qualitative by the author of \cite{16} herself. In this work, we substitute the direct contact with the soldiers by analysis of documents; the historical situation at the time was so explicit that the obtained information is sufficient. The second thought is that predictions of social phenomena are idiographic. A mechanical coupling of causes and effects which could be a base for a nomotetic regularities, even if it happens to appear, is just not interesting for social scientists. This statement is just another way of talking 
about the validity of the context.

It seems justified to say that our example is overloaded with the context. Yet, there are two paradoxical aspects here. On one hand, basically all what we know today about the Soviet policy in late 40's was clear after the famous paper by George F. Kennan, published in 1947 \cite{27}. How so many brilliant minds in West and East, not only in Poland, could trust Stalin even much later? The answer is bitter. What we want is not a prediction; the truth about our future could kill us. What we want is knowledge, how to improve our unknown future. This task needs belief and hope; both of minor intellectual reputation, but both necessary to survive hard times. On the other hand, the reader can ask what is the value of a prediction if we know that our soldiers just had no another choice? Here our answer is that it is not good to limit the research only to very difficult examples. According to Karl Popper,  `{\it the main task of the theoretical social sciences (...) is to trace the unintended social repercussions of 
intentional human actions}' \cite{1}. Then, some repercussions should be intended; as such, they have to be predictable by the wide audience.\\

Concluding, the historical process of gaining control by a prevailing power over opposing groups of people is discussed in terms of percolation theory. We have mentioned political and psychological conditions and mechanisms which made the process inevitable. In our opinion, in some countries similar mechanisms are active also today. 

\balance
\bibliography{references}
\bibliographystyle{ecms}

\end{document}